\documentclass{elsart}
\usepackage{graphicx}

\input epsf
\begin{document}
\begin{frontmatter}
\title{Bound on the anomalous $tbW$ coupling
from two-loop contribution to neutron electric dipole moment}
\author[icnunam]{A. Avilez-L\' opez},
\author[fcfmbuap]{ H. Novales-S\' anchez},
\author[fcfmbuap]{G. Tavares-Velasco},
\author[fcfmbuap]{J. J. Toscano}
\address[icnunam]{Instituto de Ciencias Nucleares, Universidad Nacional
Aut\'onoma de M\'exico, Circuito Exterior, Ciudad Universitaria,
Apartado Postal 70-543, 04510 M\'exico, D. F., M\'exico}
\address[fcfmbuap]{Facultad de
Ciencias F\'{\i}sico Matem\'aticas, Benem\'erita Universidad
Aut\'onoma de Puebla, Apartado Postal 1152, Puebla, Pue.,
M\'exico}

\begin{abstract}
The two-loop contribution to the electric dipole moment (EDM) and
the chromo electric dipole moment (CEDM) of an arbitrary fermion $f$
induced by the most general renormalizable $tbW$ coupling with
complex left- and right-handed components ($a_L$ and $a_R$) is
calculated.  The analytical expressions are numerically evaluated
and the current experimental constraints on the electron, neutron
and mercury atom EDMs are used to obtain a bound on the complex
phase ${\rm Im}(a^*_La_R)$. It is found that the most stringent
constraint, ${\rm Im}(a^*_La_R)<2.33\times 10^{-2}$, arises from the
neutron EDM.
\end{abstract}
\end{frontmatter}

Although there is experimental evidence of CP violation, its origin
still remains a mystery. In the standard model (SM), the only source
of CP violation is the Cabbibo-Kobayashi-Maskawa (CKM) phase, which
appears to be the origin of the CP violating phenomena observed in
nondiagonal processes involving the $K$ and $B$ mesons \cite{SMCPV}.
On the other hand, as diverse studies had suggested \cite{SMNDP},
the CKM phase has a rather marginal impact on flavor-diagonal
processes, such as the electric dipole moment (EDM) of elementary
particles. For instance, the EDM of fermions arises up to three
loops within the SM \cite{SMEDM}, thereby being extremely
suppressed. It can be significantly enhanced, however, in several SM
theoretical extensions, in which it can be induced at lower orders
via new sources of CP violation. It means that any experimental
signal associated with an EDM would point to new physics. Therefore,
the EDM of light fermions, such as the electron and neutron, has
been the subject of considerable interest in theories beyond the SM,
such as supersymmetric models \cite{SUSY}, multi-Higgs models
\cite{MULTIHIGGS}, left-right symmetric models \cite{LR}, and other
theories \cite{OC}. Along these lines, a potential source of CP
violation may be the $tbW$ coupling, whose study will be a top
priority at the CERN large hadron collider (LHC). Such CP-violating
effects would be induced via a complex phase arising from the
simultaneous presence of both left- and right-handed components in
the $tbW$ vertex, a scenario which is predicted indeed in several SM
extensions. The purpose of this work is to calculate the two-loop
contribution of the $tbW$ coupling to the EDM of a fermion. Although
we will obtain a result valid for any charged fermion, our main goal
is to use the experimental limits on the EDM of the electron and the
neutron to constrain the complex phase associated with the anomalous
$tbW$ vertex. As a byproduct, we will obtain the chromo electric
dipole moment (CEDM) of $f$ and use the current experimental bound
on the EDM of the mercury atom to constrain the anomalous part of
the $tbW$ vertex.

\begin{figure}[!htb]
\centering
\includegraphics[width=4.0in]{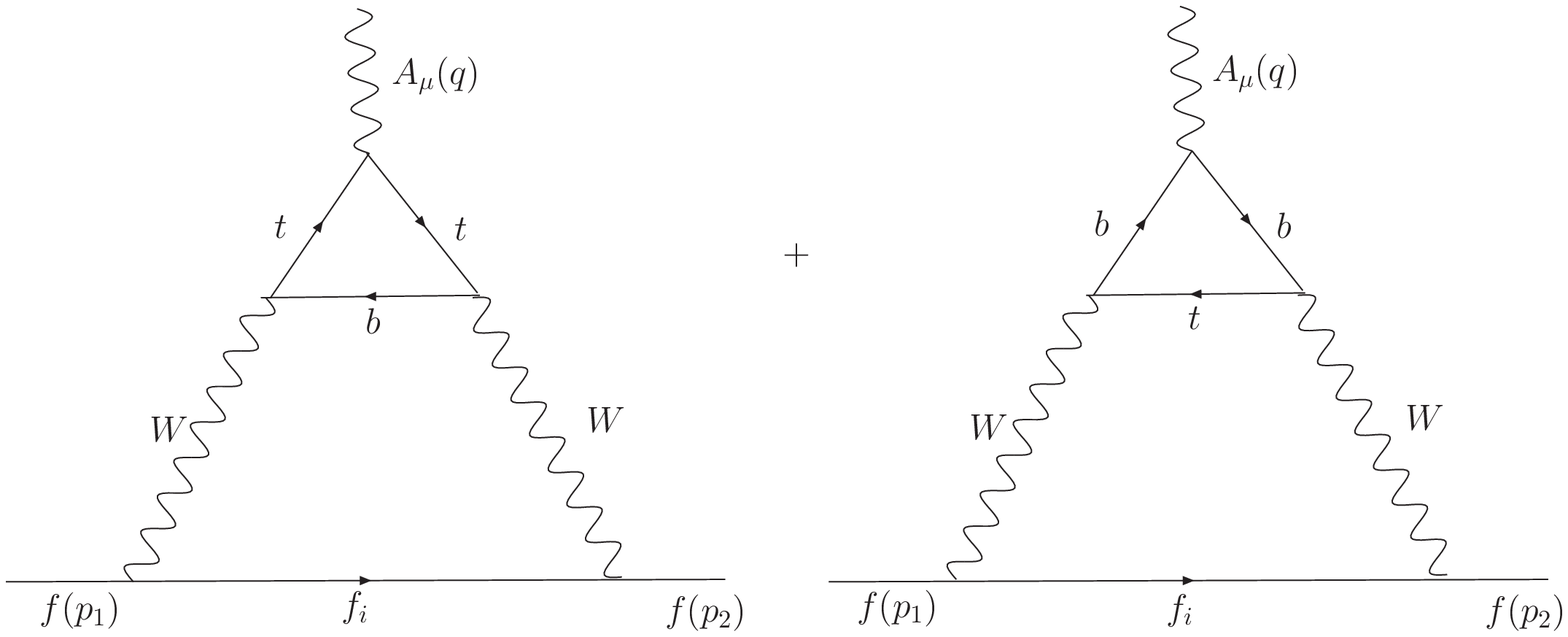}
\caption{\label{Feyndiag}Two-loop diagrams contributing to the
on-shell $\bar{f}f\gamma$ vertex. There are other diagrams in which
the photon is emitted from the $W$ boson or the internal fermion
$f_i$, but they do not contribute to the EDM of $f$. The
contribution to the CEDM arises from a similar diagram with the
photon replaced by a gluon.}
\end{figure}

The most general renormalizable  $tbW$ coupling is given by the
following Lagrangian
\begin{equation}
\label{el} \mathcal{L}=\frac{g}{\sqrt{2}}\bar{t}(a_LP_L+a_RP_R)b
W^+_\mu+{\rm H.c.},
\end{equation}
where $P_{L,R}$ are the usual left- and right-handed projectors and
$a_{L,R}$ are unknown complex coefficients. The contribution to the
on-shell $ff\gamma$ vertex, which defines the electromagnetic
properties of the fermion $f$, is given, in the unitary gauge, via
the Feynman diagrams shown in Fig. \ref{Feyndiag}. Similar diagrams,
with the photon replaced by a gluon, give rise to the CEDM. We will
present the calculation of the $ff\gamma$ vertex and then generalize
the result to the $ffg$ coupling.

From the amplitude for the on-shell $ff\gamma$ vertex we can obtain
the EDM of $f$, which is the term proportional to the Lorentz tensor
structure $i\gamma_5\sigma_{\mu \nu}q^\nu$. We will use the Feynman
parameter technique to integrate over the arbitrary internal
momenta. For the sake of completeness, we will present the most
relevant details of the two-loop calculation in appendix A.

After solving the four-dimensional integrals arising from Fig.
\ref{Feyndiag}, the EDM of $f$ can be expressed as follows

\begin{equation}
\label{df}
d_f=-\left(\frac{\alpha^2}{4\pi^2s^4_W}\right)\left(\frac{e}{2m_W}\right)
N_q x_fx_bx_t{\rm Im}(a^*_La_R)I,
\end{equation}
where $x_a=m_a/m_W$, $N_q=3$ is a quark color factor, and $I$ stands
for the quadruple integral given in appendix A [Eq. (\ref{parint})],
which must be numerically evaluated in the most general scenario,
namely, for internal and external fermions with non-negligible
masses. On the other hand, in the $x_f=x_i=0$ approximation, which
is suited for the purpose of our work as we are interested in
evaluating the EDM of very light fermions, the parametric integral
$I$ adopts the more simple form given in Eq. (\ref{integral}), which
can be simplified further after some algebra:
\begin{equation}
\label{integralsimp} I=\frac{Q_b+Q_t}{2(x_t^2-x_b^2)}+\frac{Q_t
x_b^2-Q_bx_t^2}{2(x_t^2-x_b^2)^2}\log\left(\frac{x_t^2}{x_b^2}\right)+\int_0^1
dx_1\int_0^{1-x_1}dx_2 F_3(x),
\end{equation}
with
\begin{equation}
\label{funf3}
F_3(x)=\frac{Q_b}{2}\frac{x(1-x)}{(g_b-x(1-x))^2}\log\left(\frac{g_b}{x(1-x)}\right)
-(b\leftrightarrow t).
\end{equation}

It is not possible to integrate the $F_3(x)$ function in terms of
elementary functions, but it is straightforward to obtain a
numerical solution due to the fact that $F_3(x)$ is well-behaved in
the corresponding domain.

It is useful to express the numerical value of the EDM and the the
CEDM for a light fermion. Inserting the approximate value of the
integral $I$, we obtain for a very light fermion:

\begin{equation}
\label{dgnum} |d_f|=3.194 \times 10^{-22}  x_f {\rm
Im}(a^*_La_R)\quad {\rm e\cdot cm},
\end{equation}
As for the CEDM of $f$, $\tilde d_f$, it follows easily from the
above results. We just need to make the following replacements
$\tilde d_f=d_f(e\to g_s, Q_t\to 1, Q_b\to 1)$. Numerical evaluation
gives:
\begin{equation}
\label{dggnum} |\tilde d_f|=3.056 \times 10^{-22}  x_f {\rm
Im}(a^*_La_R) \quad {\rm g_s\cdot cm},
\end{equation}

We turn to analyze our results in the light of the current
experimental limits on the EDM of fermions. Let $d_f({\rm Exp})$ be
the experimental limit of the EDM of a fermion $f$, which is not
necessarily a light one. Then, our theoretical result for $d_f$ can
be translated into the following bound on the complex phase:
\begin{equation}
\label{bound} {\rm
Im}(a^*_La_R)<\left(\frac{4\pi^2s^4_W}{\alpha^2}\right)\left(\frac{2m_W}{e}\right)
\left(\frac{m_W^3}{m_b m_t m_f}\right)\left(\frac{|d_f({\rm
Exp})|}{I}\right).
\end{equation}
This expression is exact as long as Eq. (\ref{integral}) is used for
the parametric integral $I$. In the case of a light fermion, we can
use the result of Eq. (\ref{dgnum}).

We are now ready to constrain the phase ${\rm Im}(a^*_La_R)$. For
this purpose, we will use the experimental data on the EDM of the
electron, the neutron and the mercury atom.

The current experimental limit on the EDM of the electron is
\cite{PDG}:
\begin{equation}
|d_e({\rm Exp})|< 7\times 10^{-28} \ {\rm e\cdot cm}, \\
\end{equation}
Using this limit, we obtain the following bound
\begin{equation}
{\rm Im}(a^*_La_R)<0.345.
\end{equation}

As far as the neutron is concerned, there are three different
approaches to estimate its EDM $d_n$ \cite{SMNDP}. In the chiral
Lagrangian approach, $d_n$ is expressed in terms of the quark CEDMs,
whereas in the QCD sum rule approach it is expressed as a
combination of both quark EDMs and CEDMs. Due to the large
discrepancies arising from these two approaches, it is convenient
for the purpose of this work to estimate the neutron EDM using the
non-relativistic $SU(6)$ quark model along with naive dimensional
analysis for the quark CEDM contributions. In this approach, we have

\begin{equation}
d_n = \frac{1}{3} \left (4\, d_d -d_u \right),
\end{equation}
where
\begin{equation}
\label{dq} d_{u,d} = \eta d_{u,d} + \tilde{\eta} \frac{e}{4\pi}
\,\tilde{d}_{u,d}
\end{equation}
with $\eta( \simeq 0.61)$ and $\tilde{\eta} (\simeq 3.4)$ being the
respective QCD correction factor from renormalization group
evolution. The CEDM contributions should be included as long as they
are of similar size to those from the EDMs, otherwise they can be
neglected. In our case, the quark CEDM is of similar size of the
EDM, but its contribution to Eq. (\ref{dq}) is suppressed by more of
one order of magnitude due to the factor $1/(4\pi)$. It is thus safe
to neglect the CEDM contribution. On the other hand, currently the
most stringent bound on the neutron EDM is \cite{Baker:2006ts}:
\begin{equation}
|d_n(\rm Exp)|<2.9\times 10^{-26}\ {\rm e\cdot cm}.
\end{equation}

As usual, we take $m_u\approx m_d \approx m_n/3$, with $m_n$ the
neutron mass. The above Eqs.  leads to a stronger bound than the one
found from the electron EDM:

\begin{equation}
\label{neucons} {\rm Im}(a^*_La_R)<2.33 \times 10^{-2}.
\end{equation}

As for constrains on the EDM of diamagnetic atoms, the most
stringent one can be obtained from the mercury atom. A constraint on
the ${\rm Im}(a_L^* a_R)$ can be found following the approach of
Ref. \cite{Pospelov:2001ys}, in which the experimental constraint on
the mercury EDM is translated into the bound $|\tilde{d}_u -
\tilde{d}_d| < 2 \times 10^{-26}$ cm. Explicit calculation shows
however that the resulting upper constraint on ${\rm Im}(a_L^* a_R)$
is weaker than the one obtained from the electron EDM by about one
order of magnitude. So we will not consider this constraint in this
work. We can conclude that the most stringent bound is the one
obtained from the neutron  experimental data.

We now would like to compare our constraint with other ones
appearing in the literature. It has been customary to express the
left- and right-handed parameters in the following way:

\begin{eqnarray}
a_L&=&1+\kappa_Le^{i\phi_L}, \\
a_R&=&\kappa_Re^{i\phi_R},
\end{eqnarray}
where $\kappa_{L,R}$ and $\phi_{L,R}$ are real parameters. Moreover,
$\kappa_{L,R}\geq 0$. This parametrization of the $tbW$ vertex is
simply the SM contribution ($a_L=1$, $a_R=0$) plus an anomalous
complex term expressed in polar form. It follows that our constraint
(\ref{neucons}) translates into
\begin{equation}
\kappa_R \sin\phi_R+\kappa_L\kappa_R\sin(\phi_R-\phi_L)<2.33 \times
10^{-2}.
\end{equation}
Constraints on these parameters have already been reported in the
literature. Data from $B$ meson physics allowed the authors of Ref.
\cite{GV} to impose the following limits:
\begin{eqnarray}
\label{BphysconsL}
\kappa_L\sin\phi_L&<&3\times 10^{-2}, \\
\label{BphysconsR} \kappa_R\sin\phi_R&<&10^{-3}.
\end{eqnarray}
On the other hand, the CLEO Collaboration data on the decay $b\to s
\gamma$ have been used in Ref. \cite{CPY} to constrain the
right-handed parameters:
\begin{equation}
\label{CLEOcons} -0.0035\leq \kappa_R\cos\phi_R+20 \kappa_R^2\leq
0.0039\times 10^{-3},
\end{equation}

However, as pointed out in Ref. \cite{CPY}, this constraint is not
sensitive to CP-violating effects, which are the ones we are
interested in.

It is worth combining the above constraints to find the allowed
region on the $\phi_L-\phi_R$ plane. Since Eq. (\ref{neucons})
depends on four unknown parameters, we will assume appropriate
values for $\kappa_{L,R}$. In particular, the bounds $\kappa_L\leq
0.01$ and $\kappa_R\leq 0.2$ have been derived from the $b\to
s\gamma$ decay \cite{Fujikawa:1993zu}. These bounds are somewhat
restrictive and thus larger values of these parameters are still
possible. It is thus interesting to analyze the following scenarios:
(i)$\kappa_L\sim \kappa_R$ and  (ii)$\kappa_L\gg\kappa_R$. In the
$\kappa_L\ll \kappa_R$ scenario our constraint is redundant as
(\ref{BphysconsR}) gives a tighter constraint on $\kappa_R
\sin\phi_R$. In Fig. 2 and 3 we have plotted the area on the
$\phi_L-\phi_R$ plane allowed by the constraints (\ref{neucons}) and
(\ref{BphysconsL})-(\ref{CLEOcons}) for various values of $\kappa_L$
and $\kappa_R$. The surviving region obtained after combining all
the constraints is also shown. It is easy to see that the CLEO
constraint is very constraining on the $a_R$ parameter, although it
is not useful to constrain the $\kappa_L$ parameter, let alone the
$\phi_L$ phase. The main advantage of our constraint is that it is
useful to constraint the CP violating phase and it is expected to
give a very stringent constraint on this phase once the experimental
constraint on the neutron EDM is improved.

\begin{figure}[!htb]
\centering
\includegraphics[width=4.0in]{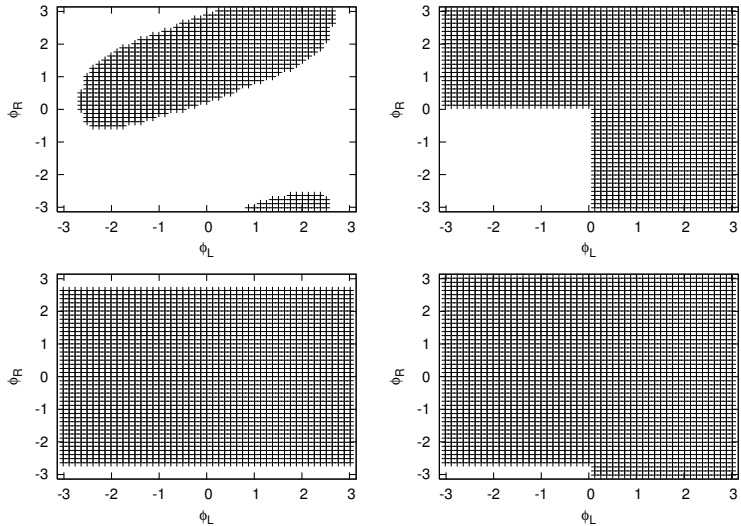}
\caption{\label{allowedarea1} Allowed area on the $\phi_L-\phi_R$
plane as obtained from the constraints (\ref{neucons}) and
(\ref{BphysconsL})-(\ref{CLEOcons}). The unshaded area represents
the area allowed by the constraint from the neutron EDM (upper
left), $B$ meson physics (upper right), the CLEO data on $b\to
s\gamma$ (lower left), as well as the surviving region obtained by
combining all the constraints (lower right). The values $\kappa_L=1$
and $\kappa_R=0.05$ have been used.}
\end{figure}

\begin{figure}[!htb]
\centering
\includegraphics[width=4.0in]{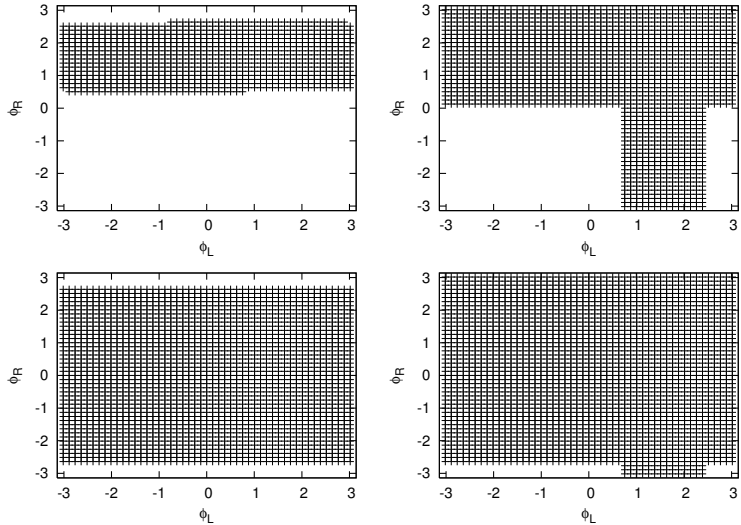}
\caption{\label{allowedarea2} The same as in Fig. 2 for
$\kappa_L=\kappa_R=0.05$.}
\end{figure}

We would like to emphasize some advantages of our method for
constraining the $tbW$ coupling. First of all, in obtaining our
constraint no extra assumptions were made. In fact, although the
result arises from a two-loop calculation, it is free of ultraviolet
divergences and the expression given in Eq. (\ref{bound}) is exact
since the parametric integral $I$ can be numerically evaluated for
any external fermion, including even a very heavy one. Our result
thus can be useful for predicting the EDM of the $\mu$ and $\tau$
leptons or the $s$ and $c$ quarks, for instance. In addition, our
bound can be easily updated. Along these lines, in Ref. \cite{PE} a
proposal was presented for improving the experimental limit of the
EDM of the electron by $3$ orders of magnitude. This would lead to a
bound on ${\rm Im} (a_L^* a_R)$ of the order of $10^{-3}$, which is
of the same order of magnitude than that arising from $B$ meson
physics. More recently, the nEDM Collaboration \cite{NC} has
presented a proposal \cite{PN} to improve the current limit on the
neutron EDM by $2$ orders of magnitude. This potential constraint
would lower the bound on ${\rm Im}(a^*_La_R)$ to the level of
$10^{-5}$, which would be more stringent by about one order of
magnitude than those obtained from $B$ meson physics and the CLEO
data.

In closing we would like to emphasize the relevance of the present
work. Important information on the origin of CP violation may be
extracted from the measurement of the EDM of elementary particles.
Several sources of CP violation are predicted in beyond-the-SM
models, and the most stringent experimental limits imposed on the
EDM of the electron or the neutron would allow us to asses their
relative importance or eventually to rule them out. In this paper,
we have studied the impact  of a complex phase associated with the
most general renormalizable $tbW$ coupling with both left- and
right-handed components on the EDM of an arbitrary $f$ fermion. For
the sake of completeness, the calculation of the two-loop amplitude
was analyzed to some extent. The resultant expression can be
straightforwardly used to predict the sensitivity of the EDM of a
light or heavy fermion to a complex phase appearing in the $tbW$
coupling. In particular, our theoretical result was numerically
evaluated in the scenario of a light fermion and the outcome was
combined with the experimental limits on the EDM of light fermions
to obtain an inequality that can be easily updated to constrain the
$tbW$ coupling. Using the most recent experimental constraints on
the EDM of the electron and the neutron, it was found that the
latter gives the most stringent bound on the $tbW$ complex phase,
which is one order of magnitude less stringent than those obtained
from $B$ meson physics. We would like to emphasize however that our
bound could be improved by about two orders of magnitude if the
neutron EDM is measured with more precision at a near future, as
recently proposed \cite{NC}.
\bigskip

{\bf{Acknowledgments.} We acknowledge financial support from
CONACYT and VIEP-BUAP (M\' exico).}
\bigskip

\appendix
\section{The two loop calculation}

Using the SM Feynman rules and the one induced by the Lagrangian
(\ref{el}), the amplitude for the $ff\gamma$ coupling can be written
as:
\begin{equation}
\mathcal{M}=\bar{u}(p_2)\Gamma_\mu u(p_1)\epsilon^\mu (q,\lambda),
\end{equation}
where $\Gamma_\mu$ is the two-loop vertex function, which is given
by
\begin{eqnarray}
\Gamma_\mu&=&\frac{ieg^4}{4}\int
\frac{d^Dk_1}{(2\pi)^D}\frac{P_R\gamma_\rho \pFMSlash{k_1}
\gamma_\lambda P^{\alpha \lambda}P^{\beta
\rho}}{[k^2_1-m^2_i][(k_1-p_1)^2-m^2_W][(k_1-p_2)^2-m^2_W]}\nonumber
\\
&&\times \int
\frac{d^Dk_2}{(2\pi)^D}\Big(\frac{Q_t}{\Delta_t}T^t_{\alpha \beta
\mu}+\frac{Q_b}{\Delta_b}T^b_{\alpha \beta \mu}\Big),
\end{eqnarray}
with $Q_t(Q_b)$ the electric charge of the $t(b)$ quark in units of
the positron charge, $m_i$ the mass of the internal fermion that
couples to the $W$ boson and the $f$ fermion, and
\begin{eqnarray}
\Delta_t&=&[k^2_2-m^2_b][(k_1+k_2-p_1)^2-m^2_t][(k_1+k_2-p_2)^2-m^2_t],\\
\Delta_b&=&[k^2_2-m^2_t][(k_1+k_2-p_1)^2-m^2_b][(k_1+k_2-p_2)^2-m^2_b],
\end{eqnarray}
\begin{eqnarray}
P^{\alpha \lambda}&=&g^{\alpha \lambda}-\frac{(k_1-p_1)^\alpha
(k_1-p_1)^\lambda }{m^2_W}, \\
P^{\beta \rho}&=&g^{\beta \rho}-\frac{(k_1-p_2)^\beta (k_1-p_2)^\rho
}{m^2_W},
\end{eqnarray}
\begin{eqnarray}
T^t_{\alpha \beta \mu}&=&Tr\Big[\gamma_\beta
(a_LP_L+a_RP_R))(\pFMSlash{k_2}+m_b)\gamma_\alpha
(a^*_LP_L+a^*_RP_R)\times\nonumber \\
&&(\pFMSlash{k_1}+\pFMSlash{k_2}-\pFMSlash{p_1}+m_t)\gamma_\mu
(\pFMSlash{k_1}+\pFMSlash{k_2}-\pFMSlash{p_2}+m_t)\Big],\\
T^b_{\alpha \beta \mu}&=&Tr\Big[\gamma_\beta (a^*_LP_L+a^*_RP_R))
(\pFMSlash{k_2}+m_t)\gamma_\alpha (a_LP_L+a_RP_R)\times\nonumber
\\
&&(\pFMSlash{k_1}+\pFMSlash{k_2}-\pFMSlash{p_1}+m_b)\gamma_\mu
(\pFMSlash{k_1}+\pFMSlash{k_2}-\pFMSlash{p_2}+m_b)\Big].
\end{eqnarray}

This amplitude generates contributions to all the form factors
associated with the on-shell $ff\gamma$ vertex, but we are only
interested in the term proportional to $\gamma_5\sigma_{\mu
\nu}q^\nu$. The latter can be isolated after the identity
$\epsilon_{\mu \nu \alpha \beta}\gamma^\alpha
\gamma^\beta=-2\gamma_5\sigma_{\mu \nu}$ is used. After some
algebra, the contribution to the EDM of $f$ can be written as
\begin{equation}
\Gamma^d_\mu=-\frac{2ie^5}{s^4_W}m_bm_t{\rm
Im}(a^*_La_R)(\gamma_5\sigma_{\mu \nu}q^\nu)\int
\frac{d^4k_1}{(2\pi)^4}\frac{\pFMSlash{k_1}}{\Delta_1}\int
\frac{d^4k_2}{(2\pi)^4}\Big(\frac{Q_t}{\Delta_t}-\frac{Q_b}{\Delta_b}\Big),
\end{equation}
where $s_W$ is the sine of the weak angle. Note that  the integrals
in $D$ dimensions were written in four dimensions as they are free
of ultraviolet divergences.

We will proceed to solve the two-loop amplitude. Using Feynman
parameters for the integral over $k_2$, one obtains
\begin{eqnarray}
\Gamma^d_\mu&=&-\frac{2ie^5}{s^4_W}m_bm_t{\rm
Im}(a^*_La_R)(\gamma_5\sigma_{\mu \nu}q^\nu)\int
\frac{d^4k_1}{(2\pi)^4}\frac{\pFMSlash{k_1}}{\Delta_1}\times
\nonumber \\
&&\Gamma (3)\int^1_0 dx_1\int^{1-x_1}_0dx_2\int
\frac{d^4k_2}{(2\pi)^4}\Bigg(\frac{Q_t}{(k^2_2-R_t)^3}-\frac{Q_b}{(k^2_2-R_b)^3}\Bigg),
\end{eqnarray}
where
\begin{eqnarray}
R_t&=&x(x-1)\Big[(k_1-l)^2-\bar{M}^2_t\Big], \\
R_b&=&x(x-1)\Big[(k_1-l)^2-\bar{M}^2_b\Big],
\end{eqnarray}
with $x=x_1+x_2$ and
\begin{eqnarray}
l&=&\frac{x_1}{x}p_1+\frac{x_2}{x}p_2,\\
\bar{M}^2_t&=&m^2_i+\frac{m^2_t}{x-1}-\frac{m^2_b}{x},\\
\bar{M}^2_b&=&m^2_i+\frac{m^2_b}{x-1}-\frac{m^2_t}{x}.
\end{eqnarray}
Once the integral over $k_2$ is done, Feynman parametrization for
the integral over $k_1$ leads to
\begin{eqnarray}
\Gamma^d_\mu&=&-\frac{2e^5}{s^4_W}m_bm_t{\rm
Im}(a^*_La_R)(\gamma_5\sigma_{\mu \nu}q^\nu)\Gamma(4)\int^1_0
dx_1\int^{1-x_1}_0 \frac{dx_2}{x(x-1)}\nonumber
\\
&&\times \int^1_0dy_1\int^{1-y_1}_0dy_2\int^{1-y}_0dy_3\int
\frac{d^4k_1}{(2\pi)^4}\pFMSlash{p}\Bigg(\frac{Q_t}{(k^2_1-\hat{M}^2_t)^4}-\frac{Q_b}{(k^2_1-\hat{M}^2_b)^4}\Bigg),
\end{eqnarray}
where we have introduced the following notation: $y=y_1+y_2$,
$p=y_1p_1+y_2p_2+ly_3$, and

\begin{equation}
\hat{M}^2_{t,b}=\bar{M}^2_{t,b}y_3+m^2_Wy+m^2_f(y+y_3)(y+y_3-1)+m^2_i(1-y-y_3).
\end{equation}

The EDM of $f$ can thus be written as

\begin{equation}
d_f=-\left(\frac{\alpha^2}{4\pi^2s^4_W}\right)\left(\frac{e}{2m_W}\right)x_fx_bx_t{\rm
Im}(a^*_La_R)I,
\end{equation}
where we have introduced the dimensionless variables $x_a=m_a/m_W$
and
\begin{equation}
I=\int^1_0dx_1\int^{1-x_1}_0dx_2\int^1_0dy_1\int^{1-y_1}_0dy_2\int^{1-y}_0dy_3
x(1-x)(y+y_3)\Bigg(\frac{Q_t}{F^2_t}- \frac{Q_b}{F^2_b}\Bigg),
\end{equation}
with
\begin{eqnarray}
F_t&=&f_ty_3+f,\\
F_b&=&f_by_3+f,\\
f_t&=&x(1-x)x^2_f+(x^2_b-x^2_t)x+x^2_b,\\
f_b&=&x(1-x)x^2_f+(x^2_t-x^2_b)x+x^2_t,\\
f&=&x(1-x)y+x(1-x)(y+y_3)(y+y_3-1)x^2_f\nonumber
\\
&&+x(1-x)(1-y-y_3)x^2_i.
\end{eqnarray}

We can further simplify the parametric integral $I$ . Explicit
integration over $y_3$ leads to
\begin{equation}
\label{parint}
I=\int^1_0dx_1\int^{1-x_1}_0dx_2\int^1_0dy_1\int^{1-y_1}_0dy_2F_1(x,y),
\end{equation}
where
\begin{eqnarray}
F_1(x,y)&=&Q_b\frac{x(1-x)}{f_b^2}\left[\left(\frac{f_b}{f}\right)\frac{(1-y)(f-f_by)}{f+f_b(1-y)}
-\log\left(1+\left(\frac{f_b}{f}\right)(1-y)\right)\right]\nonumber\\&-&(b\leftrightarrow
t).
\end{eqnarray}

It is not possible to solve the integrals over $y_1$ nor $y_2$
analytically, so numerical evaluation would be necessary. However, a
relatively simple expression can be obtained in the $x_f=x_i=0$
limit, which is a good approximation for the case of a light fermion
doublet coupling to the $W$ boson. This is suited for the purpose of
our work as we will use the experimental limits on the EDM of the
electron and the neutron to constrain the $tbW$ coupling. In such an
approximation, once the integral over $y_1$ and $y_2$ are done, the
parametric integral $I$ reduces to
\begin{equation}
\label{integral} I=\int^1_0dx_1\int^{1-x_1}_0dx_2F_2(x),
\end{equation}
where
\begin{eqnarray}
\label{funf2}
F_2(x)&=&\frac{Q_b}{2}\left[\frac{x(1-x)}{(g_b-x(1-x))^2}\log\left(\frac{g_b}{x(1-x)}\right)
-\frac{1}{g_b-x(1-x)}\right]\nonumber\\
&-&(b\leftrightarrow t),
\end{eqnarray}
with  $g_b=f_b(x_f=0)=x^2_t+(x^2_b-x^2_t)x$. The second term between
the square brackets can be integrated straightforwardly, whereas the
term with the logarithm cannot be integrated in terms of elementary
functions. The final result is given in Eq. (\ref{integralsimp}).

\end{document}